\documentclass[aps,prd,showpacs,twocolumn,preprintnumbers,amsmath,amssymb,nofootinbib,superscriptaddress,showkeys]{revtex4-1}

\usepackage{epsfig}

\begin{document}

\title{Are there three $\Xi (1950)$ states?}
 
 \author{M. Pav\'on Valderrama}\email{m.pavon.valderrama@ific.uv.es}
 \affiliation{Instituto de F\'{\i}sica Corpuscular (IFIC), Centro
 Mixto CSIC-Universidad de Valencia, Institutos de Investigaci\'on de
 Paterna, Aptd. 22085, E-46071 Valencia, Spain}

 \author{Ju-Jun Xie}
 \affiliation{Instituto de F\'{\i}sica Corpuscular (IFIC), Centro
 Mixto CSIC-Universidad de Valencia, Institutos de Investigaci\'on de
 Paterna, Aptd. 22085, E-46071 Valencia, Spain}
 \affiliation{Departamento
 de F\'{\i}sica Te\'orica, Universidad de Valencia, E-46071 Valencia, Spain}
 \affiliation{Department of Physics, Zhengzhou University, Zhengzhou,
 Henan 450001, China}

\author{J. Nieves}
\affiliation{Instituto de F\'{\i}sica Corpuscular (IFIC), Centro
 Mixto CSIC-Universidad de Valencia, Institutos de Investigaci\'on de
 Paterna, Aptd. 22085, E-46071 Valencia, Spain}

\date{\today}

\begin{abstract}
\rule{0ex}{3ex}
Different experiments on hadron spectroscopy have long
suspected the existence of several cascade states in the
$1900-2000\,{\rm MeV}$ region.
They are usually labeled under the common name of $\Xi (1950)$.
As we argue here, there are also theoretical reasons supporting 
the idea of several $\Xi (1950)$ resonances.
In particular, we propose the existence of three $\Xi(1950)$ states:
one of these states would be part of a spin-parity $\frac{1}{2}^{-}$ decuplet
and the other two probably would belong to
the $\frac{5}{2}^{+}$ and $\frac{5}{2}^{-}$ octets.
We also identify which decay channels are the more appropriate for the
detection of each of the previous states.
\end{abstract}


\maketitle

There exists scarce data on cascade ($\Xi$) resonances.
This is because (i) they can only be produced as a part of a final state,
(ii) the production cross sections are small and (iii) the final states are
topologically complicated and difficult to study with electronic
techniques. 
Thus, the bulk of information about cascade states comes
entirely from old bubble chamber experiments where the numbers of
events are small.
There are just two four star resonances, $\Xi(1318)$ and $\Xi(1530)$ with
spin--parity $J^P=\frac12^+$ and $\frac32^+$, respectively.
Those correspond to the lowest--lying $S-$wave quark model states.
Other four cascade states deserve the rating of
three stars in the Particle Data Group (PDG)~\cite{Nakamura:2010zzi}. 
Among the latter resonances, the chiral structure and spin parity of
two of them, $\Xi(1690)$ and $\Xi(1820)$, seem to be theoretically
understood~\cite{Ramos:2002xh,GarciaRecio:2003ks,Kolomeitsev:2003kt,Sarkar:2004jh,Gamermann:2011mq}\footnote{The
  $\Xi(1820)$  is dynamically generated from the
$\Delta$ decuplet--pion octet chiral interaction 
\cite{Kolomeitsev:2003kt,Sarkar:2004jh}, and it could be partner of
the $N(1520)$ in a  $\frac32^-$ octet. The
$\Xi(1690)$ and the one star $\Xi(1620)$ appear
in unitary chiral approaches to the scattering of Goldstone bosons off
baryons of the nucleon
octet~\cite{Ramos:2002xh,GarciaRecio:2003ks,Gamermann:2011mq}, and
they would be partners~\cite{GarciaRecio:2003ks,Gamermann:2011mq} of
the $N(1535), N(1650), \Lambda(1405)$ and $\Lambda (1670)$ in two
$\frac12^-$ octets.}.
The other two three star cascade resonances quoted in the PDG are
the $\Xi(1950)$ and $\Xi(2030)$ states,
which  spin-parity have not been determined yet. 
Here we will focus in these two states, in
particular in the $\Xi(1950)$ resonance.

The $\Xi (1950)$ resonance was discovered in 1965 by Badier
et al.~\cite{Badier:1965zz} in the decay channels
$K^{-} p \to \Xi^{-} K^{0} \pi^{+}$ and
$K^{-} p \to \Xi^{-} K^{+} \pi^{0}$.
The Breit-Wigner parametrization fit resulted
in a mass and width of $M = 1933 \pm 16\,{\rm MeV}$
and $\Gamma = 140\pm 35\,{\rm MeV}$ respectively.
Later, Alitti et al.~\cite{Alitti:1968zz}
confirmed the existence of a cascade resonance
with $M = 1930 \pm 20 \,{\rm MeV}$ and
$\Gamma = 80 \pm 40 \,{\rm MeV}$ in the
$K^{-} p \to \Xi^{-} \pi^{-} \pi^{+} K^{+}$ channel.
The authors of Ref.~\cite{Alitti:1968zz}
theorized that this resonance may complete the $\frac{5}{2}^{-}$ octet
composed of the $N(1675), \Lambda(1830)$ and $\Sigma(1775)$ resonances.
Several experimental searches have since then found evidence for this
state~\cite{DiBianca:1975ey,Briefel:1977bp,Biagi:1986vs,Adamovich:1999ic},
providing different and sometimes incompatible values for its mass and
width, see Table \ref{tab:experiments}.
However, the $\Xi(1950)$ has not been observed in several works
searching for $\Xi^*$ states~\cite{Hassall:1981fs,Aston:1985sn}, while
other experiments see at most a bump~\cite{Biagi:1981cu,Biagi:1986zj},
thus explaining the current three stars status for the $\Xi(1950)$ in
the PDG~\cite{Nakamura:2010zzi}.

The possibility that there may be several cascade resonances in the
$1900-2000\,{\rm MeV}$ region was first suggested by Briefel et
al.~\cite{Briefel:1977bp} who noticed that different values for the
$\Xi(1950)$ mass were to be found in different decay channels.
This expectation has been commonly discussed in later experimental
searches. Indeed,  Biagi et al.~\cite{Biagi:1981cu} commented
that {\it several bubble chamber experiments have seen indications
 of a rather broad signal in this region but in general the statistical
significance is low and it is not clear if they are all observing the
same resonance.}

\begin{table*}
\begin{center}
\begin{tabular}{|c|c|c|c|}
\hline Experiment & $M_{\Xi(1950)}$ [MeV] & $\Gamma_{\Xi(1950)}$ [MeV] & Channel \\
\hline Badier 65~\cite{Badier:1965zz} & $1933 \pm 16$ & $140 \pm 35$ & $K^{-} p \to \Xi^{-} K^0 \pi^+$, $ \Xi^{-} K^+ \pi^0$   \\
\hline Alitti 68~\cite{Alitti:1968zz} & $1930 \pm 20$ & $80 \pm 40$ & $K^{-} p \to \Xi^{-} \pi^{-} \pi^{+} K^{+}$ \\
\hline DiBianca 75~\cite{DiBianca:1975ey} & $1900 \pm 12$ & $63 \pm 78$ & $K^{-} d$ ($\Xi^{-} \pi^+$ mass distribution) \\
\hline Briefel 77~\cite{Briefel:1977bp} & $1936 \pm 22$ & $87 \pm 26$ & $K^{-} p \to \Xi^{0} \pi^{-} K^{+}$ \\
& $1961 \pm 18$ & $159 \pm 57$ & $K^{-} p \to \Xi^{-} \pi^{+} K^{0}$ \\
& $1964 \pm 10$ & $60 \pm 39$ & $K^{-} p \to \Xi(1530) \pi K$ \\
\hline Biagi 81$^*$~\cite{Biagi:1981cu} & $1937 \pm 7$ & $60 \pm 8$ & $\Xi^{-} {\rm N} \to \Xi^{-} \pi^{+} X$ \\
\hline Biagi 87a$^*$~\cite{Biagi:1986vs} & $1944 \pm 9$ & $100 \pm
31$ & $\Xi^{-} {\rm Be} \to \Xi^{-} \pi^{+} \pi^{-} X$ \\
\hline Biagi 87b~\cite{Biagi:1986vs} & $1963 \pm 5 \pm 2$ & $25 \pm 15 \pm 1$ & $\Xi^{-} {\rm Be} \to \Lambda \bar{K}^0 X$ \\
\hline Adamovich 99~\cite{Adamovich:1999ic} & $1955 \pm 6$ & $68 \pm 22$ & $\Sigma^{-} {\rm Nucleus}$ ($\Xi^{-} \pi^{+}$ mass distribution) \\
\hline
\end{tabular}
\end{center}
\caption{ Different $\Xi(1950)$ mass and width experimental
determinations. Works with an $^*$ only see a bump.} \label{tab:experiments}
\end{table*}
There are also theoretical/phenomenological reasons
to suspect for the existence of several cascade states in the vicinity
of $1950\,{\rm MeV}$. ${\rm SU}(3)$-flavor symmetry was proposed by
Gell-Mann~\cite{GellMann:1962xb} and Ne'eman~\cite{Ne'eman:1961cd} 
 as an ordering principle for hadron spectroscopy~
\cite{Samios:1974tw}. This symmetry allows to
classify baryons and mesons into multiplets of particles with the same
spin and parity. Two  consequences of ${{\rm
SU}(3)}$-flavor symmetry are the Gell-Mann-Okubo (GMO) mass
relation~\cite{GellMann:1962xb,Okubo:1961jc}, and the correlation
between the decay widths of the different hadrons conforming a
multiplet.
Here we will use the GMO mass relation to identify possible cascade resonances
with masses not far from $M = 1950\,{\rm MeV}$  and then try to match
the predicted decay widths, assuming the $\Xi(1950)$ belongs to
a particular multiplet, to the scarce experimental information available.

The GMO mass relation~\cite{GellMann:1962xb,Okubo:1961jc}
relates the masses of the baryons composing a particular multiplet.
For the octet case we have $2\left (m_{N} + m_{\Xi}\right) = 3
m_{\Lambda} + m_{\Sigma}$, while for the decuplets the GMO relation
predicts $ m_{\Omega} - m_{\Xi} = m_{\Xi} - m_{\Sigma} = m_{\Sigma} -
m_{\Delta}$.
In the fundamental octet and decuplet, these relations are satisfied
at the $1\%$ level.

We will consider three multiplets (two $\frac52^{\pm}$ octets and a
$\frac12^-$ decuplet)  for which the cascade state will possibly
lie in the $1900-2000\,{\rm MeV}$ region.
The $\frac{5}{2}^{-}$ octet would be composed of the $N(1675)$,
$\Lambda(1830)$ and $\Sigma(1775)$ resonances leading to a GMO
prediction of $m_{\Xi}[J^P=\frac52^-] = 1958 \pm 30\,{\rm MeV}$. The
error includes, added in quadratures, a $1\%$ theoretical uncertainty
for the GMO mass relation. Conversely for the $\frac{5}{2}^{+}$ octet
$\big[N(1680)$, $\Lambda(1820)$, $\Sigma(1915)\big]$ the cascade
should lie around $m_{\Xi} [J^P=\frac52^+]= 2003 \pm 24\,{\rm MeV}$.

One can firmly believe on the existence of the two cascade states
above with $J^P = \frac52^{\pm}$, since flavor SU(3) symmetry is
reasonable realized in hadron spectroscopy and the existence of
the rest of their partners is experimentally well established (four
stars in the PDG~\cite{Nakamura:2010zzi}).
These two multiplets are also derived
in~\cite{Schat:2001xr,Goity:2002pu,Goity:2003ab},
where excited baryon states were studied
in the large $N_C$ limit.

The situation is less robust in the case of the $\frac{1}{2}^{-}$ decuplet.
The existence of this multiplet is proposed in~\cite{Gamermann:2011mq}
and in the aforesaid Ref.~\cite{Schat:2001xr}.
In \cite{Gamermann:2011mq}, $J^P=\frac12^- $ $\Delta, \Sigma, \Xi$ and
$\Omega$ poles are found from a SU(6) spin--flavor
extension~\cite{GarciaRecio:2005hy} of the leading order SU(3) chiral
Weinberg-Tomozawa (WT) meson-baryon interaction.
They are grouped in a decuplet belonging to a
SU(6) spin-flavor 70 multiplet of odd parity resonances\footnote{The
WT extended interaction is strongly attractive in this spin-flavor
sector~\cite{Gamermann:2011mq}, and most of the members of the 70
SU(6) multiplet can be identified with three and four star
resonances.}, as happens in~\cite{Schat:2001xr}.
In the case of the $\frac12^- $ decuplet, there are only two known members,
the $\Delta(1620)$ and the $\Sigma(1750)$ resonances.
From the masses quoted in the PDG for them,
we estimate $m_{\Xi} [J^P=\frac12^-]= 1900 \pm 100\,{\rm MeV}$,
which is also compatible with a $\Xi$ resonance in the 1900--2000 MeV region.

According to ${\rm SU}(3)-$flavor symmetry, the decay of a baryon $a$
into a baryon $b$ and a meson $c$ takes the form~\cite{Samios:1974tw}
\begin{eqnarray}
\Gamma(a \to bc) = \frac{g^2}{8 \pi}\,
|C^a_{bc}|^2 M_b \frac{p}{M_a}\,
{\left( \frac{p}{M_s}\right)}^{2 l} \, , \label{eq:width}
\end{eqnarray}
where $g$ is a constant, $M_{a(b)}$ is the mass of baryon $a(b)$, $p$
is the center of mass momentum of the outgoing meson $c$, $l$ is the
angular momentum related with the decay and $M_s$ is a scaling mass,
which we set to $M_s = 1\,{\rm GeV}$ for simplicity. $g$ depends on
the particular multiplet assignment of the baryons $a$ and $b$. We
have only considered decays into a baryon belonging either to the
$N(940)$ octet or the $\Delta(1232)$ decuplet and a meson of the pion
octet. Thus, $g$  will stand for $g_{\mu_b}(J^{P}|_a)$, with
$\mu_b = 8, 10$ depending on whether baryon $b$ is placed in an octet
or a decuplet and $J^{P}|_a$ the spin-parity assignment of the initial
baryon $a$. Besides, $C^a_{bc}$ is the corresponding ${\rm SU}(3)$
Clebsch-Gordan coefficient, for which we follow de Swart's
convention~\cite{deSwart:1963gc}, that is, for the $8 \to 8 \otimes 8$
decays we write the coefficients in terms of the ratio\footnote{ For
instance, for the $\Xi(1950)$ partial decays we have $C(\Xi_{1950} \to
\Xi \pi) = \sqrt{3} (2\alpha - 1)$, $C(\Xi_{1950} \to \Lambda \bar K)
= \frac{4\alpha - 1}{\sqrt{3}}$, $C(\Xi_{1950} \to \Sigma \bar K) =
{\sqrt{3}}$ and $C(\Xi_{1950} \to \Xi \eta) = -\frac{2\alpha +
1}{\sqrt{3}}$.  } $\alpha = F/ (D + F)$.
\begin{table*}
\begin{center}
\begin{tabular}{|c|c|c|c|c|c|}
\hline  \multicolumn{2}{|c|}{SU(3) Decay} & Decay channel & Data $\Gamma_i$
(${\rm MeV}$) \cite{Nakamura:2010zzi} & Fitted $\Gamma_i$ (${\rm MeV}$) & Best fit parameters  \\
\hline \hline ${\frac{5}{2}}^{-}~~ 8\to8\bigotimes8$ & $l=2$ & $N(1675)\to N\pi$ &  $59 \pm 10$ & $49 \pm 7 \pm 0.0$ & $g_8=3.6 \pm 0.3 \pm 0.1$ \\
& &$\Lambda(1830)\to N\bar{K}$ & $5.5 \pm 3.4$ & $2.7 \pm 1.2 \pm 0.1$ & $\alpha = -0.23 \pm 0.06 \pm 0.00$ \\
& &$\Lambda(1830)\to \Sigma\pi$ & $47 \pm 22$ & $72 \pm 7 \pm 2$ &  $r_{g_8,\alpha}=0.75$ \\
& & $\Sigma(1775) \to N\bar{K}$ & $48 \pm 7$ & $39 \pm 5 \pm 0$ &   $\chi^2/\text{d.o.f}=1.8$\\
& & $\Sigma(1775) \to \Lambda \pi$ & $20 \pm 4$ & $26 \pm 3 \pm 0$ & \\
& & $\Sigma(1775) \to \Sigma \pi$ & $4.2 \pm 1.9$ & $3.5 \pm 1.5 \pm 0.0$ &  \\
\hline ${\frac{5}{2}}^{-}~~ 8\to10\bigotimes8$ & $l=2$& $N(1675)\to \Delta\pi$ &  $81 \pm 12$ & $86 \pm 2 \pm 0$ & $g_{10}=24 \pm 2 \pm 1$  \\
& & $\Sigma(1775) \to \Sigma(1385) \pi$ & $12 \pm 3$ & $8.5 \pm 0.4 \pm 0.1$ & $\chi^2/\text{d.o.f}=1.5$ \\
\hline \hline ${\frac{5}{2}}^{+} ~~ 8\to8\bigotimes8$ & $l=3$& $N(1680) \to N \pi$ &  $88 \pm 8$ & $81 \pm 6 \pm 0$ &$g_8=7.9 \pm 0.3 \pm 0.2$  \\
& & $\Lambda(1820) \to N \bar{K}$ &  $48 \pm 7$ & $55 \pm 6 \pm 0$ &  $\alpha=0.58 \pm 0.05 \pm 0.00$ \\
& & $\Lambda(1820) \to \Sigma \pi$ &  $8.8 \pm 2.6$ & $9.6 \pm 2.8 \pm 0.0$ &  $r_{g_8,\alpha}=-0.18$ \\
& & $\Sigma(1915) \to N \bar{K}$ &  $12 \pm 7.2$ & $3.0 \pm 2.8 \pm 0.3$ &  $\chi^2/\text{d.o.f}=1.9$ \\
\hline ${\frac{5}{2}}^{+} ~~ 8\to10\bigotimes8$ & $l=1$&  $N(1680) \to \Delta \pi$ & $13 \pm 5$ & $9.5 \pm 2.6 \pm 0.0$ & $g_{10}=2.8 \pm 0.4 \pm 0.0$ \\
& & $\Lambda(1820) \to \Sigma(1385) \pi$ &  $6.0 \pm 2.1$ & $6.8 \pm 1.9 \pm 0.0$ & $\chi^2/\text{d.o.f}=0.6$\\
\hline \hline ${\frac{1}{2}}^{-} ~~ 10 \to8\bigotimes8$  & $l=0$ & $\Delta(1620)\to N \pi$ & $36 \pm 7$ & $37 \pm 7 \pm 0.0$ & $g_{8}'=2.5 \pm 0.2 \pm 0.0$\\
& & $\Sigma(1750) \to N \bar{K}$ &  $28 \pm 21$ & $11 \pm 2 \pm 0$ & $\chi^2/\text{d.o.f}=1.1$ \\
& & $\Sigma(1750) \to \Sigma \eta$ &  $39 \pm 28$ & $5.5 \pm 1.1 \pm 3.8$ &  \\
\hline ${\frac{1}{2}}^{-} ~~ 10 \to10\bigotimes8$  & $l=2$ & $\Delta(1620)\to \Delta \pi$ &  $64 \pm 22$ & $64 \pm 22 \pm 0$ & $g_{10}'=30 \pm 5 \pm 6$ \\
\hline 
\end{tabular}
\end{center}
\caption{Experimental partial decay widths (second and third columns)
  and results from different fits of Eq.~(\ref{eq:width}): widths and
  best fit SU(3) decay parameters are displayed in the fourth and
  fifth columns, respectively. In these two latter columns, the first
  (second) set of errors stands for statistical (systematic)
  uncertainties (see text).  In the last column, we also give the
  obtained $\chi^2$/d.o.f. values for each fit, and the corresponding
  Gaussian correlation coefficients in the case of two parameter
  fits. In the first column, we give the SU(3) model details of each
  type of decays, including the value of $l$ used in
  Eq.~(\ref{eq:width}).  For the masses of the different decaying
  resonances, we have used: (i) $M = 1675 \pm 5, 1830\pm 10$ and $1775
  \pm 5\,{\rm MeV}$ for the $\frac{5}{2}^{-}$ octet,
  (ii) $M = 1685 \pm 5, 1820 \pm 5$ and $1915 \pm 20 \,{\rm MeV}$ for
  the $\frac{5}{2}^{+}$ octet and (iii) $M = 1630 \pm 30$ and $1765
  \pm 35$ MeV for the $\frac{1}{2}^{-}$ decuplet.  }
\label{tab:data}
\end{table*}

In Table \ref{tab:data} we compile experimentally known partial decay
widths of the different baryons of the $\frac52^{\pm}$ octets and
$\frac12^-$ decuplet.  Results from best fits to Eq.~(\ref{eq:width})
are shown in Table \ref{tab:data} where in addition, the values of
$\chi^2$/d.o.f., Gaussian correlation coefficients and the fitted
partial decay widths are given as well\footnote{We use a Monte Carlo
simulation to propagate the correlated errors of the fitted SU(3)
couplings, shown in the fifth column of Table~\ref{tab:data} (first
set of errors), to the partial decay widths (first set of errors in
the fourth column).}.  We observe that, given the experimental
accuracy of the data, the SU(3) flavor symmetry picture advocated here
looks consistent with data, since it provides reasonably small values
of $\chi^2$/d.o.f.  To obtain the central values and the first set of
errors in Table ~\ref{tab:data} all uncertainties in the masses have
been ignored. However, the experimental masses of the members of the
$\frac52^{\pm}$ octets and the $\frac12^-$ decuplet are certainly
poorer determined than that of each of the decay products, and one
might think that these uncertainties might have some influence both on
the determination of the SU(3) couplings and on the accuracy of the
predicted partial decay widths. To check this, we have generated
uncorrelated Monte Carlo samples for the decaying baryon masses and
have repeated the best fits for each set of mass values and
calculated, with the new fitted parameters, the corresponding partial
decay widths.  From the obtained distributions of best fit parameters
and predicted partial widths, we have read off the 68\% confidence
level intervals, which give rise to the second set of errors displayed
in Table \ref{tab:data}. In most cases, systematic errors are
much smaller than the statistical ones induced from the errors of the
decay widths used in the $\chi^2$ fits. In general,  systematic
errors are around ten times smaller than  statistical
uncertainties.

Next, we consider the $\Xi$ states of the $\frac52^{\pm}$
octets and the $\frac12^-$ decuplet, and  their SU(3)
$8\otimes 8$ and $10 \otimes 8$ decays. We use the fitted parameters
of Table \ref{tab:data} to compute the  partial
decay widths, which are shown in Table
\ref{tab:decays-1950}.  As can be seen, the
$\frac{5}{2}^{-}$ octet assignment for the $\Xi(1950)$ implies a
relatively broad resonance ($\Gamma > 100$ MeV) that should be mostly
evident in the $\Xi \pi$ invariant mass distribution. This pattern is
consistent with most of the observations of the $\Xi(1950)$,
which is usually detected in the $\Xi \pi$  channel.

On the contrary, the identification with a $\frac{5}{2}^{+}$ octet
translates into a narrow resonance visible in the $\Lambda \bar K$ and
$\Sigma \bar K$ mass distributions. These features coincide with those
of the cascade resonance found in ~\cite{Biagi:1986vs}, where a
relatively narrow cascade ($\Gamma = 25\pm 15\,{\rm MeV}$) was found
at a mass of $M = 1963 \pm 5\,{\rm MeV}$ in the $\Lambda \bar{K}^0$
mass distribution (with a statistical significance of
$3.6\,\sigma$). This experimental work was unable to find this cascade
signal in the $\Sigma^0 \bar{K}^0$ mass distribution, in apparent
contradiction with the results of Table
\ref{tab:decays-1950}. However, isospin invariance 
implies that the decay width into the $\Sigma^0 \bar{K}^0$
channel is $1/3$ of the complete decay width into $\Sigma \bar{K}$, an
observation with led the authors of Ref.~\cite{Biagi:1986vs} to the
set up the upper limit
\begin{eqnarray}
\frac{{\Gamma} (\Xi(1950) \to \Sigma {\bar K})}
{{\Gamma} (\Xi(1950) \to \Lambda {\bar K})} < 2.3 \, ,
\end{eqnarray}
at the $90 \%$ confidence level. By using the numbers of Table
\ref{tab:data}, we obtain a branching ratio of $2.2 \pm 0.6\pm 0.1$ for $M = 1965\,{\rm MeV}$, saturating (but
still compatible with) the experimental bound.
According to Ref.~\cite{Biagi:1986vs}, the spin-parity of this 
resonance should most probably be $\frac{5}{2}^{+},
\frac{7}{2}^{-}, \frac{9}{2}^{+}, \cdots$,
in agreement with our assignment.

We should comment that the $\frac{5}{2}^{+}$ identification for the
$\Xi(1963)$ state observed in \cite{Biagi:1986vs} is not entirely free
of ambiguities. Indeed, the GMO mass expectation for the
$\frac{5}{2}^{+}$ cascade, $m_{\Xi} = 2003 \pm 24\,{\rm MeV}$, looks a
bit more compatible with the $\Xi(2030)$ than with the
$\Xi(1963)$. The $\Xi(2030)$ was first observed in
\cite{Alitti:1969rb} and definitively confirmed (at the $8\,\sigma$
level) in ~\cite{Hemingway:1977uw} in the channel $K^-p \to (\Sigma
\bar K)^- K^+$. In this reference a mass $M = 2024 \pm 2\,{\rm MeV}$
and a width $\Gamma=16 \pm 5\,{\rm MeV}$ are determined, and apart
from the $\Sigma \bar K $ channel, the only other visible decay mode 
was the $\Lambda \bar K$. In fact
the PDG values, $M = 2025 \pm 5\,{\rm MeV}$ and $\Gamma=21\pm 6,{\rm MeV}$, are
mostly based on Ref.~\cite{Hemingway:1977uw}. The momentum analysis of
Ref.~\cite{Hemingway:1977uw} suggested, at the $3\,\sigma$ level, that
the spin must be $J \geq (\frac{5}{2})$ for the $\Xi(2030)$.
However, the identification of the $\Xi(2030)$ as a member of the
$\frac{5}{2}^{+}$ octet translates into a total decay width
much larger than the expected one on the basis of
Ref.~\cite{Hemingway:1977uw} ($\Gamma_{\rm th} = 76 \pm 14\,{\rm MeV}$
from Table~\ref{tab:decays-1950}
versus $\Gamma_{\rm exp} = 16 \pm 5\,{\rm MeV}$ quoted in \cite{Hemingway:1977uw}).
Ref. \cite{Hemingway:1977uw}  also determined 
\begin{eqnarray}
\frac{{\Gamma} (\Xi(2030) \to \Lambda \bar K)} {{\Gamma} (\Xi(2030) \to
\Sigma {\bar K})} = 0.22 \pm 0.09 \, ,
\end{eqnarray}
which may be incompatible with the $\Xi(2030)$ being part of the
$\frac{5}{2}^{+}$ octet, as this identification leads to the ratio
$0.4 \pm 0.1$, a $2\, \sigma$ discrepancy.
Thus, we support the identification of the $\Xi(2030)$ as
part of a different multiplet, in contrast with 
\cite{Samios:1974tw}, where this  state is assigned to be
the partner of the $\frac52^+$ $N(1680), \Lambda(1820),
\Sigma(1915)$ resonances.
\begin{table*}
\begin{center}
\begin{tabular}{|c|c|c|c|c|c|c|c|c|}
\hline 
$\Xi(1950) $ & $M$ [MeV] & $\to \Xi \pi$ & $\to \Lambda \bar K$ & $\to
\Sigma \bar K$ & $\to \Xi \eta$ & $\to \Xi(1535)\pi$ & $\to \Sigma(1385) \bar K$ & $\Gamma_{\rm Total}$ \\
\hline \hline ${\frac{5}{2}}^{-}$ octet & $1950$ &  $83 \pm 10 \pm 3$ & $14 \pm 2 \pm 1$ & $19 \pm 3 \pm 1$ & $< 0.5 $ & $19 \pm 2 \pm 1$ & $2.1 \pm 0.4 \pm 0.2$ & $137 \pm 15 \pm 5$\\
\hline \hline ${\frac{5}{2}}^{+}$ octet & $1950$ &  $1.8 \pm 1.7 \pm 0.0$ & $8.8 \pm 2.6 \pm 0.3$ & $18 \pm 1 \pm 1$ & $< 0.5 $ & $2.1 \pm 0.6 \pm 0.0$ & $0.5\pm 0.1\pm 0.0$ & $31 \pm 6 \pm 1$\\
& $1965$ &  $2.0 \pm 1.9 \pm 0.0$ & $10 \pm 3 \pm 0$ & $23 \pm 2 \pm 1$ & $< 0.5$ & $2.3 \pm 0.6 \pm 0.0$ & $0.7\pm 0.2 \pm 0.0$ & $38 \pm 7 \pm 1$\\
& $2025$ &  $3.5 \pm 3.3 \pm 0.0$ & $19 \pm 6 \pm 1$ & $48 \pm 4 \pm 2$ & $2.1 \pm 0.2 \pm 0.1$ & $3.4 \pm 0.9 \pm 0.0$ & $1.6\pm 0.4 \pm 0.0$ & $76 \pm 14 \pm 3$\\
\hline \hline ${\frac{1}{2}}^{-}$ decuplet & $1900$ &  $20 \pm 4 \pm 0$ & $17 \pm 3 \pm 0$ & $15 \pm 3 \pm 0$ & $6.8 \pm 1.4 \pm 0.1$ & $10 \pm 3 \pm 4$ & $< 0.5$ & $69 \pm 12 \pm 4$\\
& $1950$ &  $21 \pm 4 \pm 0$ & $18 \pm 4 \pm 0$ & $17 \pm 3 \pm 0$ & $11 \pm 2 \pm 0$ & $19 \pm 6 \pm 8$ & $8.0 \pm 3.0 \pm 3.3$ & $94 \pm 16 \pm 11$\\
\hline 
\end{tabular}
\end{center}
\caption{Partial and total decay widths (in ${\rm MeV}$)
of the $\Xi(1950)$ assuming it belongs to different multiplets.
The SU(3) decay parameters  are taken from Table
\ref{tab:data}.  The meaning of the two error sets given in this
table is the same as in Table \ref{tab:data}.
For the mass of the $\Xi(1950)$ we have chosen different values
in each multiplet: in the ${\frac{5}{2}}^{+}$ octet we have
considered the canonical $1950\,{\rm MeV}$ value together
with $1965\,{\rm MeV}$ (for a better comparison with Ref.~\cite{Biagi:1986vs})
and $2025\,{\rm MeV}$ (for evaluating the alternative completion of the
${\frac{5}{2}}^{+}$ octet with the $\Xi(2030)$).
In the ${\frac{1}{2}}^{-}$ decuplet we have also considered the GMO-relation
inspired value of the mass $M = 1900\,{\rm MeV}$.
} \label{tab:decays-1950}
\end{table*}

Finally, the $\frac{1}{2}^{-}$ decuplet assignment is quite unspecific
regarding the decays, see Table \ref{tab:decays-1950}.
In general this identification will lead to a broad state
that does not have a preferred decay channel.
However, if its mass is in the vicinity of $1950\,{\rm MeV}$, the
$\frac{1}{2}^{-}$ decuplet state would be the only cascade 
with a sizable $\Sigma(1385) \bar K$ branching ratio  above 5\%,
providing thus a clear signature for an eventual unambiguous
identification.

To summarize,  we have provided theoretical arguments in favor of
the experimental observation~\cite{Briefel:1977bp}
that the  $\Xi(1950)$  probably consists of several
states of similar masses.
In particular we have identified the missing cascade members of a
$\frac{1}{2}^{-}$ decuplet and the $\frac{5}{2}^{\pm}$ octets as
possible candidates for explaining different appearances of the
$\Xi(1950)$.
While the $\frac{1}{2}^{-}$ decuplet signal would be quite indistinct,
the $\frac{5}{2}^{-}$ octet identification fits into the experimental
observations of broad structures in the $\Xi \pi$ invariant mass
distribution (e.g. the old Ref.~\cite{Badier:1965zz} or the more
recent work of Ref.~\cite{Adamovich:1999ic}), while the
$\frac{5}{2}^{+}$ assignment is compatible with the observation of a
narrower state in the $\Lambda \bar K$ decay channel and with a mass
of about 1965 MeV~\cite{Biagi:1986vs}.
We disfavor, however, the identification~\cite{Samios:1974tw} of
the $\Xi(2030)$ as the missing member of the $\frac52^+$ octet.
We find it worth mentioning
that Refs.~\cite{Schat:2001xr,Goity:2002pu,Goity:2003ab}
lead to the same multiplet assignments as this work
from a different theoretical background,
and suggest, in addition, the existence of
even more cascades in the $1900-2000\,{\rm MeV}$ region.

\begin{acknowledgments}

This research was supported by DGI and FEDER funds, under contracts
FIS2006-03438, FIS2008-01143/FIS, and the Spanish
Consolider-Ingenio 2010 Programme CPAN (CSD2007-00042),  by Generalitat
Valenciana under contract PROMETEO/2009/0090, by the EU
HadronPhysics2 project, grant agreement n. 227431 and by the
National Natural Science Foundation of China (NSFC) under grant
n. 11105126. 

\end{acknowledgments}

\end{document}